\newcounter{myctr}

\documentclass{ws-acs}
\usepackage{amssymb,amsmath}
\begin{document}

\makeatletter
\def\@biblabel#1{[#1]}
\makeatother

\markboth{O. Cabrera, D. H. Zanette}{Avoiding extinction by
migration: The case of  the head louse}

%
\catchline{}{}{}{}{}
%

\title{AVOIDING EXTINCTION BY MIGRATION: \\
THE CASE OF THE HEAD LOUSE}

\author{OCTAVIO CABRERA}

\author{\footnotesize DAMI\'AN H. ZANETTE \footnote{Also at Consejo
Nacional de Investigaciones Cient\'{\i}ficas y T\'ecnicas, Argentina.}}

\address{Centro At\'omico Bariloche
and Instituto Balseiro \\ 8400 San Carlos de Bariloche, R\'{\i}o
Negro, Argentina \\
cabreraoct@gmail.com, zanette@cab.cnea.gov.ar}

\maketitle

\begin{history}
\received{(received date)}
\revised{(revised date)}
\end{history}

\begin{abstract}
The possibility of spreading by migration, colonizing new spatial
domains suitable for development and reproduction, can substantially
relieve a biological population from the risk of extinction. By
means of a realistic computational model based on empirical data, we
study this phenomenon for the human head louse, {\it Pediculus
humanus capitis}. In particular, we show that a lice colony
infesting a single isolated host is prone to extinction by
stochastic population fluctuations within an interval of several
months, while migration over a relatively small group of hosts in
contact with each other is enough to insure the prevalence of the
infestation for indefinitely long periods. We characterize the
interplay of the size of the host group with the host--to--host
contagion probability, which controls a transition between
extinction of the lice population and a situation where the
infestation is endemic.
\end{abstract}


\section{Introduction}
\label{sect1}

The evolution of the number of individuals in a large, spatially
homogeneous biological population can be well approximated by smooth
functions of time, governed by a system of ordinary differential
equations. The trend of the population to flourish or decline
results from the balance of average birth, death,  and migration
rates which, ordinarily,  are in turn smoothly time--varying
quantities \cite{Murray1}. Spatial inhomogeneity can readily be
incorporated to this kind of continuous description by means of
partial differential equations \cite{Murray2}.

On the other hand, small populations are likely to undergo sharp
variations in size even over short time spans, due to the effect of
random fluctuations \cite{fluct}. For instance, a rapid succession
of deaths of fertile individuals may lead to a bottleneck  that
severely impairs growth of the population in the short term,  and
perhaps induce  its extinction. The smaller the population,
moreover, the more pronounced the effect is: under homogenous
environmental conditions and taking into account stochastic death
events only, the probability that all the members of a population of
$N$ individuals  die within a given period is $p_N= p_1^N$, where
$p_1$ is the death probability of a single individual during the
same time. As a consequence, for example, if a group of animals
becomes extinct with a probability of $0.0001$ during the
non--breeding  season, a similar group of half its size becomes
extinct  with a  probability  of $0.01$. The extinction of the
second group is, therefore,  a hundred times  more likely.

A significant illustration of the risk of extinction that threatens
small populations is given by the fact that the main criterion to
define endangered animal and plant species is, precisely, their
reduced number of individuals \cite{extinction}. Another major
criterion of extinction risk regards  the limited extension and high
fragmentation of their geographical distribution. In contrast, under
favorable conditions, the capability of  expanding the distribution
by colonizing new regions may guarantee access to the resources
necessary for the population to grow in size,  thus decreasing the
exposure to extinction.  Even a geographically fragmented population
can avoid local extinction by timely migration between different
habitat patches \cite{PRL}. If, nevertheless, extinction in a patch
does occur,  it can be later recolonized by individuals of the same
species  coming from  other zones.

In this paper, we address the possibility of avoiding population
extinction by  colonization of new spatial domains in the specific
case of the human head louse, {\it Pediculus humanus capitis}.
Recent gathering of empirical results on the louse's life parameters
made it possible to put forward realistic mathematical models for
the demography of lice colonies \cite{LR}, on which we base our
analysis. In a single host's head, the number of lice at any given
moment is limited to, at most, a few tens of individuals. The host's
self--grooming, which becomes more intense as the number of lice
grows, controls the proliferation of nymphs and adults. Moreover,
the number of eggs laid by each fertilized female during its whole
lifespan is relatively small. As a result, as we show here, the
average waiting time for a single--host lice colony to become
extinct by random fluctuations ---disregarding any kind of specific
treatment--- is typically within the order of a  few to  several
months. The possibility of migrating from one host to another, thus
establishing new colonies, is therefore crucial for the survival of
the infestation over much longer time spans.

After formulating our computational model on the basis of empirical
observations of the louse's life cycle, we first present extensive
numerical simulations for single--host colonies, characterizing
their survival probability and average duration as functions of the
parameters associated to the host's grooming. Then, we move to the
case of several hosts, where we find our main result: for a fixed
rate of migration between hosts, there is a minimum number of hosts
above which the lice population survives indefinitely, the
infestation thus becoming endemic. Conversely, for a  group of hosts
of a given size, the infestation becomes endemic if the migration
rate is large enough. The transition between extinction   and
endemic prevalence is characterized numerically. Moreover, we show
that it can be qualitatively explained by means of a simple
metapopulation model. Conclusions and perspectives are drawn in the
last section.

\section{The life of lice: Empirical data and computational model}

Lice are parasitic insects, belonging to the genus {\it Pediculus}
among others, that infest almost every bird and mammal species.
Pediculosis, the infestation of lice, has been a steadfast companion
of mankind until powerful insecticides began to be used against it
by the mid twentieth century.  The ensuing decrease in the
prevalence of pediculosis led head--louse infestation, its most
common variety, to be regarded as a signature of lack of hygiene and
precarious life conditions. In the developed world, pediculosis thus
became a kind of social taboo, which also resulted into a relative
loss of interest from  health scientists. Systematic studies of the
biology and epidemiology of lice  had to wait a few decades, until
pediculosis regained prevalence ---probably due to the combined
effect of lice's developed resistance to insecticides and the
banishment of the most aggressive chemicals from treatment
\cite{reg}.

In the last decade, a few controlled studies have quantitatively
characterized the life cycle of the louse \cite{TL,LR7}.  These
results have recently been used to formulate a mathematical model
for the population dynamics of a lice colony \cite{LR}. The data
relevant to our contribution can be summarized as in the following.

\subsection{Life history of the head louse} \label{life}

As for many other insects, the life of a louse spans three stages:
egg, nymph, and adult. The nymph, in turn, undergoes three moults
before becoming an adult. Observations of {\it in vivo} and {\it in
vitro} lice colonies show that some $76 \pm 3$\% of the eggs are
viable and bear a nymph, each of them hatching $8.4 \pm 0.1$ days
after it was laid. Recorded durations of the egg stage vary between
$7$ and $11$ days. The duration of the nymph stage is even better
defined, with recorded values of $8$ and $9$ days, depending on sex
and strain \cite{TL}.

In cultivated colonies, some $23$\% of the nymphs die before
reaching adulthood. The adult stage, in turn, lasts $12$ days on the
average, with a maximal recorded duration of $23$ days \cite{TL}.
Under natural life on a host, mortality rates grow due to the host's
self--grooming, mainly as combing, washing and scratching. Grooming,
on the other hand, does not affect eggs, which are firmly glued to
the hair. Both nymphs and adults are mobile and feed on their host's
blood several times a day \cite{LR7}. Consequently, they survive
only a few hours if they are taken away from the host.

After the first or second day of adulthood, each fertile female lays
$4.9 \pm 0.2$ eggs per day \cite{TL}. Scattered studies starting at
the nineteenth century suggest a female--biased sex ratio of around
$60$\%  \cite{sexr}. Plausibly, a single adult male suffices to
fertilize all the adult females in a head's colony which, under
normal conditions, consists of $10$ to $20$ mobile (nymph plus
adult) individuals. It has been moreover suggested that, once
fertilized, a female remains fertile for its whole life.

Pediculosis mostly affects children, from the age of $3$ to the
beginning of puberty. Hundreds of millions of children worldwide are
affected each year. Females seem to be more prone to get infested.
In any given social group of children ---for instance, at school or
in sport activities--- the typical prevalence ranges between $2$ and
$20$\%  \cite{reg}. It is apparent from direct observation of such
groups that contagion plays a crucial role in the dynamics of head
lice infestations.  Transmission from head to head occurs by direct
motion of the lice when the hosts are close enough to each other, or
by transportation on inanimate objects such as combs, clothes, and
toys \cite{LR10,LR11}.  Symptoms of pediculosis ---mainly, itching
caused by allergic reaction to the lice bites--- appear when the
number of mobile individuals reaches a certain threshold, which has
been estimated to be around $15$ lice \cite{LR}. This triggers more
intense grooming and, consequently,  nymph and adult mortality
increases.

Quantitative results on the transmission dynamics of head lice in
realistic conditions ---making it possible, for instance, to assign
numerical values to host--to--host contagion probabilities--  seem
not to be not available yet \cite{stoch,TL2005,pelo}. It is
conceivable, moreover, that such parameters are strongly dependent
on specific circumstances of the host group, ranging from
socioeconomic conditions and cultural traits to the kind of social
activity they carry on together. The effectiveness of grooming below
and above the awareness threshold, in turn, has not been measured.
Similarly, ovicidal treatments ---both  chemical and with lice
combs--- are difficult to evaluate during actual use \cite{LR38}.
This kind of treatment, in any case, is not considered in the
present  contribution.

\subsection{Computational agent--based model} \label{cmodel}

On the basis of the data presented above, we formulate a
computational model for an evolving population of lice, according to
the rules described below. Time elapses in discrete steps, each step
corresponding to a day. Each louse is represented by an agent
endowed with several attributes. An ``internal clock,'' updated at
every time step, measures the louse's age in days, starting at the
day where the egg was laid by the mother. At any time, the louse's
life stage (egg, nymph, or adult) is determined from its age, using
the parameters reported below. A binary index defines the louse's
sex (female or male). Another index records whether an adult female
is fertilized or not. Finally, in populations distributed over
several hosts, an additional index identifies the louse's present
host.

In our model, as explained below, only viable eggs ---namely, eggs
which successfully bring forth a nymph--- are taken into account.
Hatching takes place for all eggs at day 9. The nymph stage, in
turn, lasts 8 days, thus ending at day 17. During this stage, there
is a constant spontaneous death probability of $0.03$  per day.
Under this sole effect, on the average, the nymph population reduces
to a fraction of approximately $0.78$ during the whole stage. As
explained below, however, this mortality must be increased with the
contribution from death events caused by the host's grooming.

During the adult stage, the probability of spontaneous death of a
louse grows to $0.083$ per day. If lice would be allowed to reach
arbitrarily old ages, this would determine an average duration of
$12$ days for adults. For computational reasons, however, we limit
the duration of this stage to a maximum of $23$ days, so that all
surviving lice die exactly $40$ days after the corresponding eggs
were laid.

Fertilization of a  female requires that, at any day of its adult
life, it shares its host with an adult male. Once this happens, the
female remains fertilized for the remaining of its life. Fertilized
females can lay eggs after the second day as adults. Every day, each
one of these females lays $5$ eggs. The viability probability of
each egg is fixed at $0.76$ so that, on the average, $3.8 \pm 2.1$
viable eggs per fertilized female per day are incorporated to the
population. Nonviable eggs are discarded. The sex is determined
probabilistically at the moment in which each egg is laid, with a
female--biased sex ratio of 60\%.

We model the effect of the host's grooming as an addition to the
death rate of nymphs and adults, assuming that  ordinary grooming
takes place below a certain awareness threshold, whereas above the
threshold  ---when symptoms of the infestation appear--- more
intense grooming is triggered. When the total number of mobile lice
in the host's head is less than a critical value $n_c$, their
spontaneous death probability per day is increased by $0.05$, so
that it reaches $0.08$ for nymphs and $0.133$ for adults. When, on
the other hand, the number of mobile lice grows beyond the threshold
$n_c$, death probabilities are added a further contribution $p_c$.
Both $n_c$ and $p_c$ remain as free parameters, as we aim at
studying how results on single--host infestation change when they
are varied.

Finally, for populations distributed over several hosts, it is
necessary to fix the per--day probability $p_s$ that a mobile louse
passes from one host to another. This ``jump'' probability is also
kept as a free parameter.

In all cases, the initial condition for the population consists of a
single fertilized female in the third day of adulthood, inhabiting
one of the hosts. At every time step in the computational
realization of the model, the population is subjected to the above
rules, according to their respective conditions, in the following
order: (1) unfertilized adult females become fertilized, (2) newly
laid eggs are added to the population, (3) the ``internal clock'' of
all agents is updated, (4) agents are allowed to move from one host
to another, and (5) death events are applied and dead agents
removed.

By comparison with the mathematical model on which the present
computational scheme is based \cite{LR}, we are here being
parsimonious with respect to the number of parameters, in
particular, regarding the durations of the egg and nymph stages.
Indeed, instead of admitting that the durations are statistically
distributed ---which requires to define the respective probability
distributions--- we assume that they are deterministically defined.
This simplification, however, seems to be reasonably justified by
empirical observations of the narrow variation of those durations
(see Section \ref{life}). In our model, thus, the dispersion in the
life span of individual lice and the increasing overlapping of
successive generations is respectively controlled by mortality
during the nymph and adult stages, and by randomness in the times of
egg laying.

\section{Single--host colonies} \label{singleh}

In this section, we study the population dynamics of a lice colony
infesting a single host. The results set the basis for the analysis
of populations on several hosts, when contagion is allowed for.

We first consider a lice population where the effect of the
awareness threshold is disregarded. Death rates, hence, are those
determined by the probability of spontaneous death of nymphs and
adults plus the contribution of sub--threshold grooming. Although
there is a small probability that the initial female dies without
having laid any viable egg, thus terminating the colony during the
first few days, the typical situation is that a growing population
does develop. Since, in the absence of the awareness threshold, the
number of birth and death events is on the average proportional to
the number of lice in different life stages, it is expected that the
population grows exponentially in time.

\begin{figure}[th]
\centerline{\psfig{file=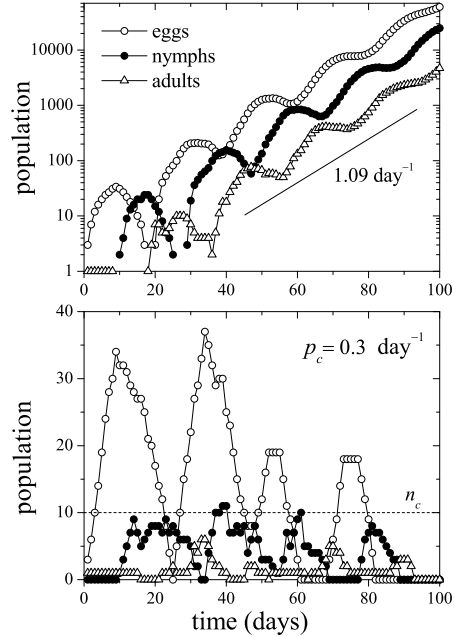,width=6cm}} \vspace*{8pt}
\caption{Upper panel: Number of eggs, nymphs, and adults as a
function of time (log--linear scales) in a single computational
realization for a lice population where the host's awareness
threshold is disregarded. The straight line stands for the average
slope of the exponential growth. Lower panel: As in the upper panel
(linear--linear scales), for a population with an awareness
threshold of $n_c=10$ mobile lice, and an additional death
probability due to grooming above threshold fixed at $p_c=0.3$
day$^{-1}$. The population becomes extinct by the death of the last
adult at day $93$.} \label{fig1}
\end{figure}

The upper panel of Fig.~\ref{fig1} illustrates this situation in a
single realization of our computational model. Different symbols
represent the recorded number of lice in each life stage as a
function of time. In this log--linear plot, straight lines with
positive slope correspond to exponential growth. We find that the
three sub--populations asymptotically approach this kind of growth
through damped oscillations. The period of these oscillations is
approximately given by the age at which a fertilized adult female is
first able to lay eggs, namely, $19$ days after its own egg was
laid. Oscillation damping is due to the effect of fluctuations in
the duration of successive generations along each lineage, as the
population progresses. Note that, asymptotically, the  numbers of
eggs, nymphs and adults represent fixed fractions of the total
population, respectively, around 67\%, 28\%, and 5\%. As indicated
by the straight line in the figure, the average growth slope of the
population equals $1.09$ day$^{-1}$. In other words, the population
increases by $9$\% every day. This is in reasonable quantitative
agreement with other mathematical models  for lice populations
\cite{LR} which, under similar conditions, predict a growth rate of
$12$ to $13$\% per day.

Naturally, no human host is expected to endure in their head a lice
colony that attains hundreds or thousands of individuals. The above
results show, however, that ---for the chosen set of parameter
values--- the trend of the lice population below the awareness
threshold is to grow with time. The effect of introducing the
threshold $n_c$, with an additional, sufficiently large death
probability $p_c$, is precisely to inhibit the unlimited growth of
the population. The lower panel of Fig.~\ref{fig1} shows an example
for a single realization with $n_c=10$ and $p_c=0.3$ day$^{-1}$
(note the linear--linear scales). Initially, due to the succession
of life stages, oscillations similar to those in the upper panel
take place in the three sub--populations. Now, however, the number
of mobile lice varies irregularly around the threshold. Since the
sub--populations of nymphs and adults never attain very large sizes,
fluctuations in the frequency of birth and death events eventually
lead to the extinction of the whole colony.

\begin{figure}[th]
\centerline{\psfig{file=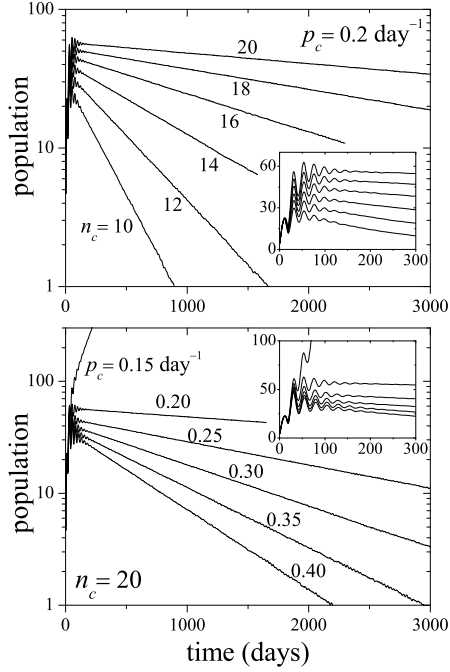,width=6cm}} \vspace*{8pt}
\caption{Total population ---eggs plus nymphs plus adults--- as a
function of time,  averaged over $10^5$ realizations of the lice
colony, for $p_c=0.2$ day$^{-1}$ and several values of   the
awareness threshold $n_c$ (upper panel), and for $n_c=20$ and
several values of $p_c$ (lower panel). The insets show close--ups
for short times. Note that the main plots have log--linear scales,
while in the insets they are linear--linear.} \label{fig01}
\end{figure}

In order to provide a statistical description of the time evolution
of the population, we have performed series of $10^5$ realizations
for each parameter set, averaging the number of lice over
realizations at each time step up to a maximum of $3000$ days
($\approx 8$ years, the typical maximal duration of pediculosis
during childhood). Figure \ref{fig01} shows the average total
population (eggs plus mobile individuals) as a function of time, for
several choices of the parameters $n_c$ and $p_c$. We find that,
after an initial stage with damped oscillations of a period of about
$20$ days ---displayed in more detail in the insets--- the average
population settles down to an essentially monotonic time dependence,
either increasing (for $n_c=20$, $p_c=0.15$; lower panel) or
decreasing. This long--time behaviour is well  approximated by an
exponential  (a straight line in the log--linear  scales of the
plot).

The fact that, for fixed $n_c$ and depending on the value of $p_c$,
the average total population grows or shrinks with time points to
the  existence of a critical value for $p_c$ above which the lice
colony eventually becomes extinct. Analyzing in detail intermediate
values of $p_c$, we have found that the critical probability is $p_c
= 0.17 \pm 0.01$ day$^{-1}$ for awareness thresholds varying from
$n_c=10$ to $20$.  This critical value coincides with the estimation
given in previous work \cite{LR}. As for its dependence on $n_c$, we
expect that for unrealistically small and large values of the
awareness threshold, $p_c$ becomes respectively lower and higher. In
fact, for small $n_c$, even very small values of $p_c$ would lead to
a rapid extinction of the colony due to fluctuations. On the other
hand, large values of $n_c$ would permit the accumulation of many
eggs even if the death probability of mobile lice above the
threshold is very large.

The exponential decay of the average total population (for $p_c$
above the critical value) reveals that its long--time evolution
would be well described by a linear equation, of the type $\dot n =
-\alpha n$ with $\alpha >0$. As expected, our results show that the
extinction rate $\alpha$ grows as $n_c$ decreases and $p_c$
increases. From the epidemiological viewpoint, however,  it is
important to provide a statistical quantification of the duration of
the lice colony that includes both the long--time behavior and the
initial stages of the evolution. To this end, we define the survival
probability $P(t)$ as the probability that a population starting at
$t=0$ is still extant at time $t$, i.e. that its whole duration is
equal or larger than $t$. The survival probability can be written as
\begin{equation}
P(t)= \sum_{u=t+1}^\infty p(u),
\end{equation}
where $p(u)$ is the probability that the population becomes extinct
exactly  at time $u$. Note that the average duration, $ T =
\sum_{u=0}^\infty u p(u)$, has a simple expression in terms of the
survival probability, as
\begin{equation} \label{avdur}
T = \sum_{t=0}^{\infty} P(t).
\end{equation}
We have evaluated $P(t)$ from our series of $10^5$ realizations, for
which we recorded the total duration of the population. The fraction
of realizations where the duration is equal to $u$ gives a direct
estimate of $p(u)$.  Realizations where the lice population was
still extant at the maximum  time of  $3000$ days were counted  and
added to $P(t)$ for all $t\le 3000$.

\begin{figure}[th]
\centerline{\psfig{file=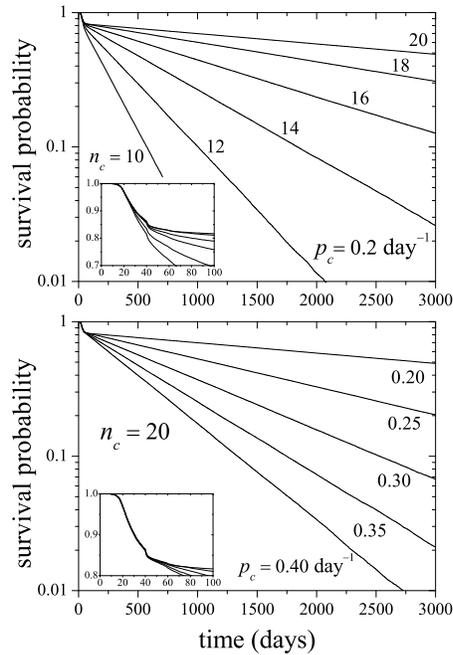,width=6cm}} \vspace*{8pt}
\caption{Upper panel:  The survival probability $P(t)$ as a function
of time for $p_c=0.2$ day$^{-1}$ and several values of the awareness
threshold $n_c$, calculated from series of $10^5$ realizations for
each parameter set (log--linear scales). The inset shows a close--up
for short times (linear--linear scales). Lower panel:  As in  the
upper panel, for $n_c=20$ and several values of $p_c$.} \label{fig2}
\end{figure}

The upper panel of Fig.~\ref{fig2} shows the survival probability
$P(t)$ as a function of time in log--linear scales, for  $p_c=0.2$
day$^{-1}$ and several values of the awareness threshold $n_c$. The
inset is a close--up for short times. Like for the average total
population as a function of time (Fig.~\ref{fig01}), we find that
the survival probability exhibits two well defined regimes. For
short times ($t\lesssim 30$ days, or a month), $P(t)$ is independent
of the threshold $n_c$. In this regime, the survival probability is
controlled by the spontaneous extinction of the first generation of
lice, having never attained a sufficiently large number of mobile
individuals. At the end of this period, some $10$\% of the colonies
have become extinct. For long times ($t> 100$ days, or about three
months), on the other hand, an exponential tail develops. As
expected, the slope is less steep as $n_c$ increases and longer
durations become more likely. In the intermediate zone ($30 \lesssim
t \lesssim 100$), where the curves for different $n_c$ begin to
diverge from each other, the wavy profile of $P(t)$ reveals the
effect of the oscillations referred to in connection with the
results shown in the lower panel of Fig.~\ref{fig1}, which make
extinction more probable when the initial generations conclude their
life cycles.

Similar results are shown in the lower panel of Fig.~\ref{fig2}, for
a fixed awareness threshold $n_c=20$ and several values of $p_c$.
Now, the short time regime where the survival probability is
independent of $p_c$ extends up to $t \approx 45$ days, where
slightly above $15$\% of the colonies have become extinct. For long
times, the exponential tails are steeper as $p_c$ increases.

\begin{figure}[th]
\centerline{\psfig{file=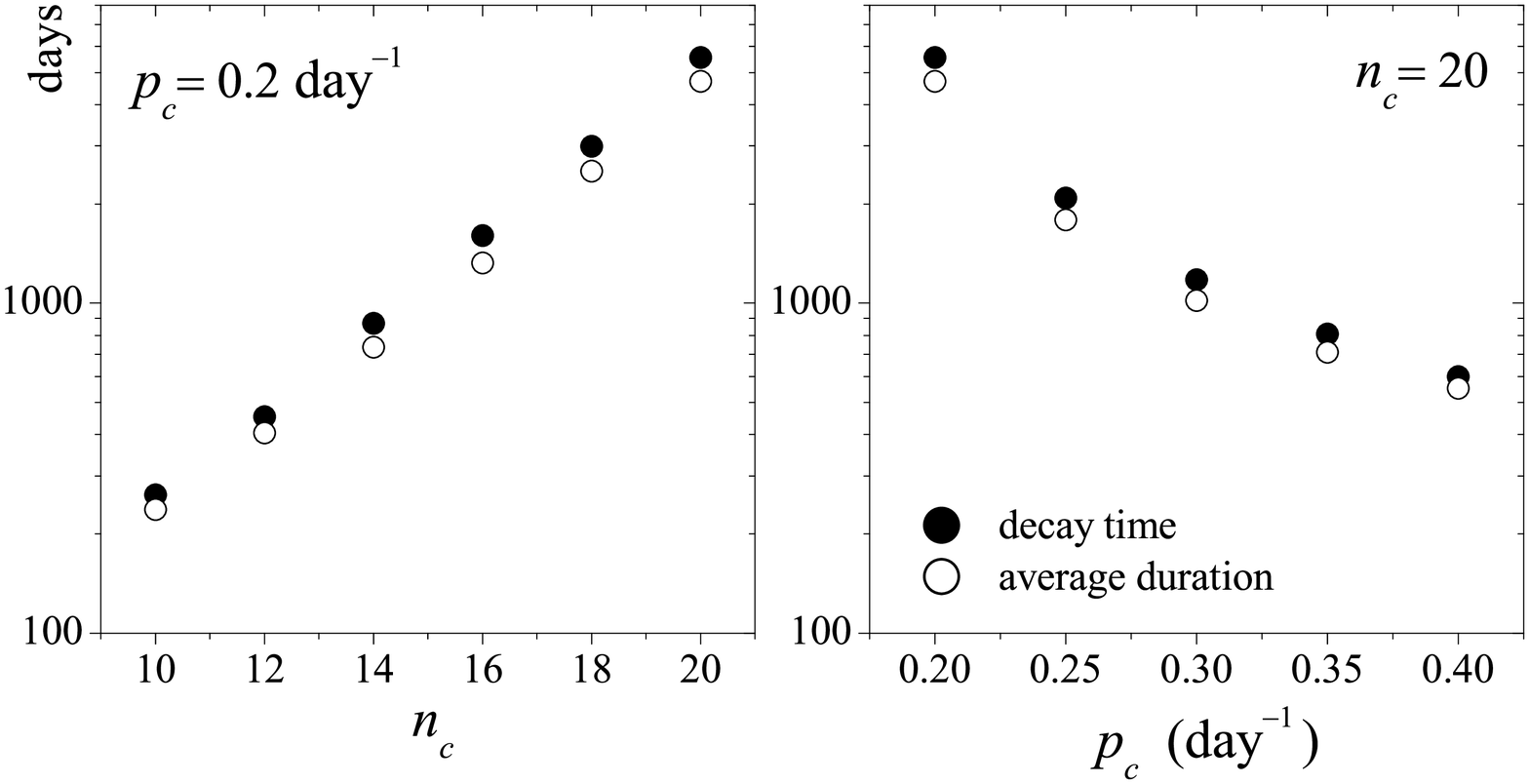,width=8cm}} \vspace*{8pt}
\caption{Left panel: Decay time of the exponential tail of the
survival probability ($\tau$, full dots),  and average duration  of
colonies ($T$, empty dots) as functions of $n_c$ for $p_c=0.2$
day$^{-1}$, calculated from series of $10^5$ realizations for each
parameter set. Right panel: As in the left panel, for $n_c=20$ and
as functions of $p_c$.} \label{fig3}
\end{figure}

The  fact that $P(t)$ decays exponentially for long times implies
that, similarly, the probability $p(u)$ that the duration of a
colony equals $u$ is exponential for large $u$: $p(u) \sim \exp
(-u/\tau)$. The characteristic decay time $\tau$ is the same for
$p(u)$ as for $P(t)$. Results of the computation of $\tau$ from
least--square  fitting of the data displayed in Fig.~\ref{fig2} are
shown in Fig.~\ref{fig3} as full dots.  Moreover, we have evaluated
the average colony duration $T$ using Eq.~(\ref{avdur}), obtaining
the empty dots of   Fig.~\ref{fig3}. Note that, for each parameter
set, the values of $\tau$ and $T$ are very similar to each other,
with the former systematically above the latter. As a matter of
fact, if $P(t)$ were a purely exponential function,  $\tau$ and $T$
would exactly coincide. Due however to the deviations for short
times (shown in the insets of Fig.~\ref{fig2}), which imply a
relatively high frequency of short durations, the average duration
is shorter than the decay time.  The difference is slightly larger
for large $n_c$ and small $p_c$, where the decay times are longer.

\section{Lice populations on several hosts}

On the basis of the results obtained in the preceding section, we
now move on to consider a lice population infesting a group of
several hosts. As advanced in Section \ref{cmodel}, contagion
between hosts is described by the per--day probability $p_s$ that a
louse ``jumps'' from  one host to another. We disregard any
underlying structure in the host group, so that contagion is equally
probable between any pair of hosts. We recall that our main aim is
to evaluate how the number of hosts $H$ affects the possibility that
lice colonies survive by extending its spatial distribution.

\begin{figure}[th]
\centerline{\psfig{file=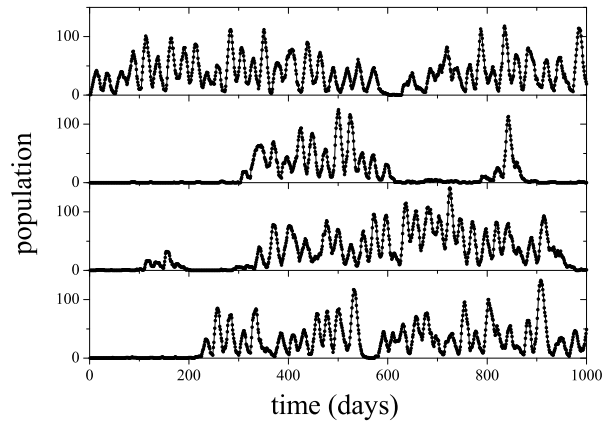,width=8cm}} \vspace*{8pt}
\caption{Individual lice populations in a four--host group, with a
jump probability between hosts $p_s=0.03$ day$^{-1}$. With this
probability, each louse moves from its present host to any of the
others. The awareness threshold is $n_c=12$, and the additional
death probability due to grooming above the threshold is $p_c=0.2$
day$^{-1}$.} \label{fig5}
\end{figure}

Each plot in Fig.~\ref{fig5} represents the lice population as a
function of time in a single host's head, out of group of $H=4$
hosts, with a jump probability $p_s=0.03$ day$^{-1}$.  The awareness
threshold is $n_c=12$ and the additional death probability due to
grooming above the threshold is $p_c=0.2$ day$^{-1}$. Initially, the
infestation is confined to the host represented in the uppermost
panel. At later stages, it is transmitted by contagion to the other
three hosts. Note that, occasionally, the population in an
individual host may fall to zero, and  reappear later.

Rather straightforward arguments make it possible to predict that,
under conditions where a single--host colony becomes extinct, mutual
contagion between a sufficiently large group of hosts may however
insure the survival of the lice population at arbitrary long times,
and make the infestation endemic.  Using for instance Levins
time--continuous model for metapopulation dynamics \cite{metap}, the
number $h(t)$ of  infested hosts, out of a group of $H$ hosts, is
expected to vary  as
\begin{equation} \label{Levins}
\dot h = -\frac{h}{\tau} + p_h h (H-h).
 \end{equation}
Here, $\tau$ is the average duration of a single--host colony (under
the assumption that the colony's survival probability decays
exponentially with time; see Section \ref{singleh}), and $p_h$ is
the host--to--host contagion probability per time unit. The
contribution of contagion is proportional to the number of infested
and non--infested hosts, implying logistic evolution for $h(t)$.
According to this model, $h(t)$ either decays to $h_\infty=0$ or
approaches a finite value $h_\infty = H - (\tau p_h )^{-1}$ at
asymptotically long times, depending on whether $H$ is lower or
larger than $H_c=(\tau p_h )^{-1}$, respectively. In other words,
whereas for $H<H_c$ the whole lice metapopulation  becomes extinct,
for $H>H_c$ it persists indefinitely, infesting a finite number of
hosts. Note the reciprocal relation between the critical number of
hosts and the contagion probability $p_h$. Moreover, for $H$ both
above and below $H_c$ and for sufficiently long times, the number of
infested hosts approaches its asymptotic value $h_\infty$ as
$|h(t)-h_\infty | \sim \exp (-s t)$, with $s=|H p_h-\tau^{-1}|$. The
linear dependence of the exponential slope $s$ with the parameters
$H$ and $p_h$ is a signature of the transcritical nature of the
transition between extinction and survival in Levins model
\cite{transcritical}.

In connection with our description ---where lice colonies are not
treated as metapopulations, but as multiagent populations--- the
host--to--host contagion probability in Eq.~(\ref{Levins})  must be
associated with the jump probability $p_s$. Meanwhile, as we have
seen in  Section \ref{singleh}, the average duration of a
single--host colony is determined by the whole set of parameters
that define the louse's life history and the host's grooming.
Whereas we do not foresee Levins model to reproduce our results at a
quantitatively reliable level, we do expect to find a transition
between extinction and survival of the lice population as the number
of hosts $H$ grows,  including the intuitive inverse relation
between the critical number of hosts and the jump probability.

\begin{figure}[th]
\centerline{\psfig{file=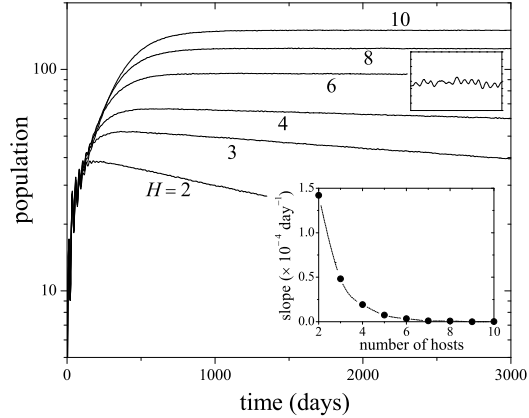,width=7cm}} \vspace*{8pt}
\caption{Mean per--host population as a function of time,  averaged
over $10^5$ realizations, for  $n_c=12$, $p_c=0.2$ day$^{-1}$, and
jump probability $p_s=0.01$ day$^{-1}$, on groups of $H=2$ to $10$
hosts. The upper inset is a close--up of the curve for $H=6$ for
times between $2500$ and $2800$, amplifying by factors of $2$ and
$15$ in the horizontal and vertical axes, respectively. The lower
inset shows the asymptotic exponential slope $s$, obtained from
least--square fittings for long times. The dotted curve is a spline
interpolation for guiding the eye.} \label{fig6}
\end{figure}

Figure \ref{fig6} displays the mean lice population per host as a
function of time, on groups of $2$ to $10$ hosts, with jump
probability $p_s=0.01$ day$^{-1}$. As we see from Figs.~\ref{fig01}
(upper panel) and \ref{fig3} (left panel), the parameters $n_c=12$
and $p_c=0.2$ day$^{-1}$ correspond to a situation where a
single--host colony becomes extinct within an average time of a few
hundred days. For larger host groups, in contrast,  Fig.~\ref{fig6}
shows that the long--time population decay becomes slower as the
number of hosts increases. For $H \gtrsim 8$, moreover, no decay can
be perceived within the time range of the plot.  The population
seems  to asymptotically reach a stationary level, which grows as
the host group increases in size.  In qualitative terms, this is
precisely the behavior predicted by Levins model.

Quantitative differences are apparent, however, when we measure the
exponential slope of the population decay as a function of the
number of hosts. Results are displayed in the lower inset of
Fig.~\ref{fig6}.  The slope decreases and approaches zero with a
very smooth dependence on the number of hosts,  suggesting  a
higher--order  transition from extinction to survival.  This
smoothness jeopardizes the precision with which the transition point
could be determined. Moreover, as discussed in more detail below,
the slopes themselves are calculated up to a  precision defined by
the small irregular fluctuations of the population averaged over
realizations, visible in the close--up  of Fig.~\ref{fig6}. The size
of these fluctuations is controlled by the number of realizations
($10^5$ in our case), and does not decrease as the number of hosts
increases. This fact amounts to larger relative errors in the
determination of smaller slopes,  as $H$ grows. Finally, we cannot
count upon improving our description of the transition by such
standard methods as finite--size analysis as, being a system driven
by the fluctuations of the population, the dynamics is not scalable
to other  population sizes or host numbers \cite{finitesize}.

In view of these drawbacks, we analyze the transition from
extinction to survival by an alternate method that,  rather than
focusing on the determination of critical points and exponents,
seeks to characterize how the system approaches the transition as
the control parameters are varied. Still, the method is based on the
analysis of the average time dependence of the lice population, as
shown by the curves of Fig.~\ref{fig6}.  For each of these curves,
we define a normalized long--time slope as $S=(n_{\max}-
n_{\min})/n_{\max}$, where $n_{\max}$ is the maximum of the mean
per--host population averaged over $10^5$ realizations, and
$n_{\min}$ is the minimum reached by the same quantity in the time
interval between the maximum and $t=3000$.

\begin{figure}[th]
\centerline{\psfig{file=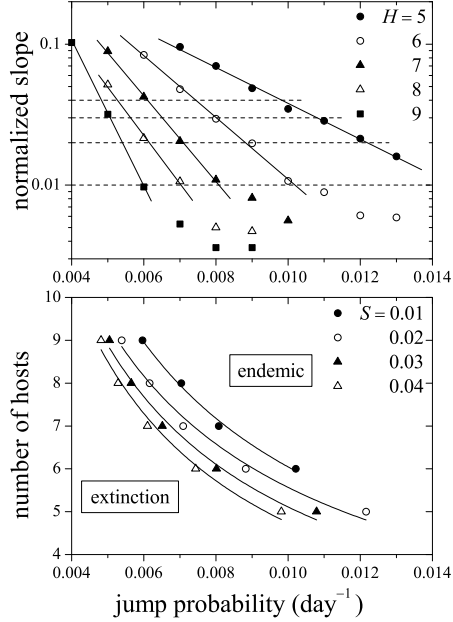,width=6cm}} \vspace*{8pt}
\caption{Upper panel: Symbols stand for the normalized slope $S$ as
a function of the jump probability $p_s$ for several values of the
number of hosts $H$, calculated for the mean per--host lice
population as a function of time, averaged over $10^5$ realizations.
The awareness threshold is $n_c=12$ and the death probability above
the threshold is $p_c=0.2$ day$^{-1}$. Full straight lines are
exponential least--square fittings above $S=0.01$, for each value of
$H$. Horizontal dashed lines are a set of selected levels of $S$,
used to calculate the data shown in the lower panel. Lower panel:
Symbols stand for the intersections between the selected levels of
$S$ and the exponential fittings for each value of $H$, plotted in
the ($p_s,H$)--plane. Curves are algebraic fittings of the form $H
\propto p_s^{-\gamma}$. The exponent $\gamma$ ranges in the interval
($0.7,0.9$) for the four curves.} \label{fig7}
\end{figure}

Symbols in the  upper panel of Fig.~\ref{fig7} stand for our
determination of $S$ for host groups of size $H=5$ to $9$ and
several values of the jump probability $p_s$.  As in
Fig.~\ref{fig6}, we have taken $n_c=12$ and $p_c=0.2$ day$^{-1}$.
These results suggest that, for $S \gtrsim 0.01$, the normalized
slope decreases exponentially with $p_s$, with steeper decays for
larger values of $H$.  Full straight lines in the plot correspond to
least--square exponential fittings for each value of $H$,  including
all the data with $S\ge 0.01$. The $R$--coefficients of the fittings
are in all cases larger than $0.99$. Below $S=0.01$, clearly, the
exponential dependence breaks down, and the normalized slope seems
to approach a constant value as $p_s$ grows further.  It is not
difficult to realize that this regime results from the small
fluctuations in the population as a function of time, caused by
averaging over realizations, which are the dominant effect in the
determination of $S$ when the curves become almost horizontal.
Plausibly, the threshold could be lowered by averaging over more
realizations, but the same effect would ultimately control the
resulting value of $S$ for sufficiently small slopes. Our analysis,
henceforth, focuses on the data above the threshold.

Once the exponential fittings of $S$ as a function of $p_s$ have
been determined for each value of $H$, we choose a decreasing
sequence of levels for the normalized slope, $S=0.04$, $0.03$,
$0.02$, and $0.01$, indicated by the horizontal dashed lines in the
upper panel of Fig.~\ref{fig7}. For each of these levels, we
analytically determine its intersection with each exponential
fitting, thus defining a pair ($p_s,H$). Each pair gives an
estimation for the jump probability and the number of hosts that, on
the average, are expected to give a long--time dependence of the
lice population which decays with the corresponding slope $S$. The
pairs ($p_s,H$) determined by this procedure are plotted in the
lower panel of Fig.~\ref{fig7}.

Inspired by Levins model, which predicts for the transition point an
interdependence of the form $H \propto p_h^{-1}$ between the number
of hosts and the host--to--host contagion probability, we fit to the
pairs ($p_s,H$) an algebraic function of the form $H \propto
p_s^{-\gamma}$ for each value of $S$. The resulting exponents are
all in the range $\gamma  = 0.8 \pm 0.1$, with a slight tendency to
decrease for small values of $S$. The $R$--coefficients of the
fittings are above $0.98$. While, from this analysis, we are not
able to define a critical transition line in the parameter plane
($p_s,H$), we can safely associate the region above the curves with
the endemic regime of pediculosis, while the lower region
corresponds to the extinction of the lice population.

Finally, we recall that the above numerical results for lice
populations on several hosts hold for the choice $n_c=12$ and
$p_c=0.2$  day$^{-1}$. A simple dimensionality argument makes it
possible to conjecture how the results can be extended to different
values of the parameters that characterize grooming above the
threshold. In fact, the interdependence between the
(non--dimensional) number of hosts $H$ and the jump probability
$p_s$ (whose units are day$^{-1}$) must involve an additional
quantity with units of time, in order to cancel the units of $p_s$.
In the long--time stage of the evolution, this quantity is the decay
time $\tau$ of the one--host survival probability, introduced in
Section \ref{singleh}.  This decay time fixes the scale of the
demographic processes whose competition with host--to--host
contagion determines the transition between extinction and the
endemic state. In turn, $\tau$ is determined by $n_c$, $p_c$, and
the life--cycle parameters of lice. Note that the result obtained
from Levins metapopulation equation for the critical number of
hosts, $H_c=(\tau p_h )^{-1}$, agrees with this argument. In our
model, it amounts to conjecture that the algebraic relation assumed
to hold between $H$ and $p_s$ for constant $S$, and illustrated in
the lower panel of Fig.~\ref{fig7}, should be written as $H \propto
(\tau p_s)^{-\gamma}$. The rather strong dependence of $\tau$ on
both $n_c$ and $n_p$ (see Fig.~\ref{fig3}) suggests in turn that
small variations in any of these parameters may have important
effects in the transition point. For instance,  for $p_c=0.2$
day$^{-1}$, changing the awareness threshold from $n_c=12$ to $14$
duplicates the value of $\tau$. This implies that, for a given
number of hosts, the same long--time dynamics is obtained with half
the jump probability.

\section{Conclusion}

Colonies of {\it Pediculus humanus capitis}, the head louse,
typically consist of a relatively small number  of individuals on
each human host. This is due to the limiting intervention of the
hosts themselves who, usually, do not tolerate the symptoms of
pediculosis without actively contributing to lice mortality through
grooming or other treatments. On a single host, this maintains the
colony on the verge of extinction, likely to disappear by mere
stochastic fluctuations in its population. Moreover, lice die within
a few hours if they are removed from their hosts and cannot have a
timely meal of blood. The continued prevalence of pediculosis along
mankind's history ---inherited, in turn, from other primates--- must
thus be explained taking into account the possibility of contagion
from host to host. The head louse thus constitutes a remarkable
example of survival by expansion of the spatial distribution:
migration between different domains ---the hosts' heads--- makes it
possible that the population, otherwise jeopardized by  aggressive
external factors, grows and becomes established.

To quantitatively characterize this phenomenon, we have here used a
computational model based on realistic empirical data from the
louse's life cycle. The model is an adaptation of a recently
proposed mathematical formulation used to evaluate the effect of
various treatment strategies for  pediculosis \cite{LR}. We have
shown that, on the average, a lice colony infesting a single host
never exceeds some tens  of individuals, and takes a time of the
order of several months to disappear by population fluctuations.
When contagion between hosts becomes possible, on the other hand, a
group of less than ten hosts in mutual contact already makes it
possible that the lice population survives indefinitely. These
quantitative results, however, have required to assign values to
certain parameters for which there are no systematic empirical
observations ---namely, the host's awareness threshold, the effects
of  grooming on lice mortality, and the contagion probability.
Although  our choices sound to be reasonably realistic, at least in
their order of magnitude, more precise data  would help to  sharpen
the present conclusions.

It is interesting to remark that, despite of the complex
fluctuation--control\-led demographic dynamics of each individual
colony, the average behavior of a lice population distributed over
several hosts is well described ---at least, qualitatively--- by a
simple metapopulation model which only takes into account the mean
life of a colony on a single host and the contagion probability
between hosts. This model predicts a reciprocal relation between the
critical number of hosts necessary for the infestation to reach an
endemic state and the contagion probability. In our more detailed
description, the possibility of contagion is implicitly (but not
directly) determined by the probability that each individual louse
passes from one host to another.  In spite of this difference, we
also find an inverse relation between the number of hosts necessary
to yield a given overall decay in the lice population and this
``jump'' probability, although the relation involves a different
exponent. This suggests that, as in other models for distributed
demography \cite{PRL}, the transition between extinction and the
endemic state is ultimately controlled by a straightforward
competition between the time scales of local population decay, on
one side, and migration, on the other.

In our analysis, we have not considered any mechanism able to remove
eggs from the hosts' hair.  In this sense, we have been describing
host groups with no access to pharmacological treatments whose use
against pediculosis is standard in the present day ---or groups that
purposely avoid their use.  The extension of the model to include
such mechanisms is in principle self--evident,  though it would
require a quantitative evaluation of their efficacy.  Taking into
account that, on the average, some two thirds of the lice population
are normally in the egg stage,  effective removal of eggs should
translate into a substantial decrease of the mean lifetime of a
colony \cite{LR}.  According to our results,  this decrease makes a
much larger host group necessary for the infestation to become
endemic. The study of groups formed by many hosts, in turn, may
require taking into account the structure and dynamics of social
interactions inside each group, and their interplay with the
epidemiology of pediculosis, an interesting topic toward which our
work calls for being extended.

\section*{Acknowledgments}
We thank Drs.~Fabiana Laguna and  Sebasti\'an Risau Gusm\'an for
their critical reading of the manuscript. Financial support from
ANPCyT (PICT 2011--545) and SECTyP UNCuyo (Project 06/C403),
Argentina, is acknowledged.


\begin{thebibliography}{00}

\bibitem{Murray1}
J. D. Murray, {\it Mathematical Biology I. An Introduction}
(Springer, Berlin, 2002).

\bibitem{Murray2}
J. D. Murray, {\it Mathematical Biology II. Spatial Models and
Biomedical Applications} (Springer, Berlin, 2003).

\bibitem{fluct}
R. Lande, S. Engen, and B.--E. Saether, {\it Stochastic Population
Dynamics in Ecology and Conservation} (Oxford University Press,
2003).

\bibitem{extinction}
G. M. Mace, N. J. Collar, K. J. Gaston, C. Hilton--Taylor, H. R.
Ak\c{c}akaya, N. Leader--Williams, E. J. Milner--Gulland, and S. N.
Stuart, Quantification of extinction risk: IUCN's system for
classifying threatened species, Conserv. Biol. {\bf  22} (2008)
1424--1442.

\bibitem{PRL}
M. Khasin,  B. Meerson,  E. Khain, and L. M. Sander, Minimizing the
population extinction risk by migration, Phys. Rev. Lett {\bf 109},
138104 (2012).

\bibitem{LR}
M. F. Laguna and S. Risau--Gusm\'an, Of lice and math: Using models
to understand and control populations of head lice, PLoS ONE {\bf 6}
(2011)  e21848.

\bibitem{reg}
N. Gratz, {\it Human lice, their prevalence and resistance to
insecticides} (World Health Organization, Geneva, 1998).

\bibitem{TL}
M. Takano--Lee, K. S. Yoon, J. D. Edman, B. A. Mullens, and J. M.
Clark, In vivo and in vitro rearing of {\it Pediculus humanus
capitis} (Anoplura: Pediculidae), J. Med. Entomol. {\bf 40} (2003)
628--635.

\bibitem{LR7}
R. Speare, D. V. Canyon, and W. Melrose, Quantification of blood
intake of the head louse: {\it Pediculus humanus capitis}, Int. J.
Dermat. {\bf 45}  (2006) 543--546.

\bibitem{sexr}
M. A. Perotti, S. S. Catal\'a, A. Orme\~no, M.  {\.Z}elazowska, S.
M. Bili{\'n}ski, and H. R. Braig, The sex ratio distortion in the
human head louse is conserved over time, BMC Genetics {\bf 5} (2004)
10.

\bibitem{LR10}
C. N. Burkhart and C. G. Burkhart, Fomite transmission in head lice,
J. Amer. Acad. Dermat. {\bf 56} (2007) 1044--1047.

\bibitem{LR11}
D. V. Canyon and R. Speare, Indirect transmission of head lice via
inanimate objects, Open Dermat. J. {\bf 4} (2010) 72--76.

\bibitem{stoch}
P. Stone, H. Wilkinson--Herbots, V. Isham, A stochastic model for
head lice infections, J. Math. Biol. {\bf 56} (2008) 743--763.

\bibitem{TL2005}
M. Takano--Lee, J. D. Edman, B. A. Mullens, and J. M. Clark,
Transmission potential of the human head louse,  {\it Pediculus
capitis} (Anoplura: Pediculidae), Int. J. Dermat.  {\bf 44}
 (2005) 811--816.

\bibitem{pelo}
D. V. Canyon, R. Speare, and R. Muller, Spatial and kinetic factors
for the transfer of head lice ({\it Pediculus capitis}) between
hairs, J. Invest. Dermat. {\bf 119}  (2002) 629--631.

\bibitem{LR38}
S. Sonnberg, F. A. Oliveira, I. L. Ara\'ujo de Melo, M. M. de Melo
Soares, H. Becher, J. Heukelbach, {\it Ex vivo} development of eggs
from head lice ({\it Pediculus humanus capitis}), Open Dermat. J.
{\bf 4} (2010) 82--89.

\bibitem{metap}
I. Hanski, Single--species metapopulation dynamics,  Bio. J. Linn.
Soc. {\bf 42} (1991)  17--38.

\bibitem{transcritical}
P. G. Drazin, {\it Nonlinear Systems} (Cambridge University Press,
Cambridge, 1994).

\bibitem{finitesize}
K. Binder and D. W. Heermann, {\it Montecarlo Simulation in
Statistical Physics} (Springer, Berlin, 1988).

\end{thebibliography}
\end{document}